\documentclass[slac_one]{revtex4}
\usepackage{graphicx,axodraw}
\usepackage{fancyhdr}
\def\as{\alpha_s}
\def\Q{{\mathcal{Q}}}
\def\D{{ D}^{\rm ini}}

\begin{document}

\title{
\begin{flushright}
\vbox{
\begin{tabular}{l}
\small UH-511-1078-05\\
\small  DESY 05-212\\
\   \end{tabular} }
\end{flushright}
The Heavy Quark Fragmentation Function at NNLO
\footnote{Based on talks given at the Linear Collider meeting
LCWS05 at Stanford University in April 2005, at the Workshop on
LHC/ILC Synergies at SLAC in April 2005 and at the DIS 2005
Workshop in Wisconsin, Madison in April 2005.}}

\author{Alexander Mitov}
\affiliation{University of Hawaii, Honolulu, HI-96822, USA\\
and\\
Deutsches Elektronensynchrotron DESY, Platanenallee 6, D-15735
Zeuthen, Germany}

\begin{abstract}

We present a general discussion of collider processes with
not-completely inclusive production of heavy flavors. We review
the Perturbative Fragmentation Functions formalism as the
appropriate tool for studying such processes and detail the
extension of this formalism at Next-to-Next-to-Leading order. We
conclude that the prospects for the future in this field are
bright and at present are limited by the available experimental
data. Hopefully, the future ILC will be able to fill this gap.

\end{abstract}

\maketitle
\thispagestyle{fancy}

\section{INTRODUCTION: PRODUCTION OF HEAVY FLAVORS}

Production of heavy flavors presents special challenges to theory.
There are two reasons for that. The first one is technical and is
related to the appearance of an additional parameter (the quark
mass) in the integrals. This is a complication that could lead to
a manifold increase in the computational effort. The second reason
is, however, more subtle and is the subject of the present
discussion.

Let us consider some not-completely inclusive differential
observable. A typical example would be a one particle inclusive
cross-section like the energy spectrum of a quark $q$ in the
reaction $e^+e^-\to q\overline{q}$. If the observed parton is
massless the differential distribution is divergent once the
radiative corrections to the process are included. These collinear
divergences arise from configurations where two or more partons
become collinear. To deal with the collinear singularities one
first introduces a suitable regulator. The most convenient and
popular method is the dimensional regularization which
simultaneously regularizes all UV divergences. After the usual UV
renormalization is performed, one arrives at the following result
for the desired massless differential cross-section:
\begin{equation}
{d\sigma_a \over dz}(z,Q,\epsilon) = \sum_b {d\hat\sigma_b \over
dz}(z,Q,\mu)\otimes\Gamma_{ba}(z,\mu,\epsilon), \label{facN0}
\end{equation}
where $b$ runs over the relevant quark flavors and the gluon, and
$\otimes$ stands for the usual integral convolution.

Eq.(\ref{facN0}) demonstrates that the collinear divergences in
{\it any} differential distribution completely factorize, i.e.
they are solely contained in the function $\Gamma$. Of course, the
explicit form of this function depends on the regularization
method chosen and on the convention for the non-divergent pieces
that $\Gamma$ contains. It has become standard practise to
factorize the collinear divergences in the so-called
$\overline{\mathrm{MS}}$ scheme. The most important property of
the collinear "counterterm" $\Gamma$ is its universality, i.e. its
form is independent of the process $d\sigma$. The physical meaning
of the collinear divergences appearing in differential
distributions like Eq.(\ref{facN0}) is well understood; they are
related to long distance effects and should be absorbed into the
fragmentation function of the experimentally observed hadronic
state.

The simple picture described above does not work if the produced
quark is massive (a relevant example is the process $e^+e^-\to
b\overline{b}$). Naively, one observes that the corresponding
energy distribution is no longer divergent. The finiteness of the
result, however, has little to do with its applicability. A closer
inspection shows that the spectrum contains terms of the type $
\sim \ln(Q/m)$ at all orders in the perturbative expansion. Here,
$Q$ stands for the typical hard scale of the process and $m$ is
the mass of the ``observed' final quark. For example, in
$e^+e^-\to b\overline{b}$, the scale $Q$ is identified with the
center of mass energy $\sqrt{s}$ and $m$ with the pole mass of the
$b$-quark.

In QCD, the six quark flavors are typically divided in light and
heavy according to the size of their mass relative to
$\Lambda_{\rm QCD}$: if $m\lesssim \Lambda_{\rm QCD}$ then the
flavor is light, while for $m\gg \Lambda_{\rm QCD}$ it is heavy.
In applications $c$,$b$ and $t$ quarks are considered heavy. The
relevant for our discussion reason for such separation is the fact
that light and heavy quarks hadronize very differently: the former
hadronize at scales of the order of the typical hadronic scale,
while the later hadronize at a scale set by the mass of the heavy
quark. Clearly, at a scale of the order of the mass of the heavy
quark perturbation theory is still valid since $\alpha_s(m)$ is
sufficiently small to be within the perturbative domain.

At the same time it is intuitively clear that heavy flavors can
behave as massless in reactions where the typical hard scale $Q$
is much larger than the mass $Q>>m$. To illustrate the point, one
can imagine taking formally the limit $m/Q\to 0$; then the
corresponding differential distribution diverges due to
logarithmic terms $\sim\ln(m/Q)$, i.e. we get situation similar to
the one in the pure massless case discussed above but with the
collinear divergences regulated with small quark mass rather than
dimensionally.

It is now easy to understand the reasons leading to the above
mentioned complications in the case of heavy quark production: one
cannot neglect the quark mass altogether since it sets the very
important hadronization scale. Yet, in presence of very large hard
scales, the mass becomes ``irrelevant" and plays merely the role
of a regulator for the collinear divergences. In such situation
one may wish to find a way to compute hard cross-sections with
massless quarks (because it is simpler) and at the same time to
keep the relevant information on the mass in the differential
distribution.

\section{THE PERTURBATIVE FRAGMENTATION FUNCTION}

The method that combines the convenience of massless,
dimensionally regularized calculations while keeping the relevant
information about the mass of the heavy quark was proposed by Mele
and Nason \cite{MN} and is known as the Perturbative Fragmentation
Function (PFF) formalism. The method relies on the factorization
of long and short distance effects in QCD by making use of the
DGLAP evolution equation to resum the above mentioned large logs
$\sim \ln^k(m/Q)$ to all orders in the strong coupling constant.

Before we describe in detail how the PFF method works, lets us
explain the need for resummation of the quasi-collinear logs. In
practical applications one can never achieve the strict limit
$m/Q\to 0$ since the hard scale $Q$ is always ``finite", i.e. we
work with finite quantities at each perturbative level. The large
(though finite) difference between $m$ and $Q$ shows up in a
somewhat disguised way - through the convergence of the
perturbative expansion. Even if $\alpha_s(m)$ is sufficiently
small and one can expect reasonable convergence of the
perturbative series, the presence of large number like $\ln(m/Q)$
at each order of the coupling effectively alters the expansion
parameter from $\alpha_s$ to $\ln^k(m/Q)\alpha_s$, $k\geq 0$,
which may not be small anymore. To circumvent this problem one has
to sum up the whole perturbation series.

The resummation of classes of terms $\sim \ln^k(m/Q)\alpha_s^n$ is
possible. This follows from the fact that the logs
$\sim\ln^k(m/Q)$ are of universal collinear origin and can all be
predicted with the help of the QCD evolution equations.

Let us consider the production of a heavy quark of flavor $\Q$
with mass $m$ and a definite value of energy $E_\Q$ in a hard
scattering process. According to the QCD factorization theorems
the heavy quark energy spectrum can be computed as a convolution
of the energy distribution of massless partons produced in the
hard process, and the fragmentation function that describes the
probability that the massless parton fragments into a massive
quark with a definite energy. If the energy fraction $E_\Q/E_{\Q ,
max}$ of the heavy quark is denoted by $z$, then the energy
distribution of that quark can be written as:
\begin{equation}
{d\sigma_\Q \over dz}(z,Q,m) = \sum_a\int_z^1{dx\over x}
{d\hat{\sigma}_a \over
dx}(x,Q,\mu){D}_{a/\Q}\left(\frac{z}{x},\frac{\mu}{m} \right) +
{\cal O}\left({m\over Q}\right)^p. \label{fac}
\end{equation}
Here, the sum runs over all partons (i.e. quarks, antiquarks and
gluons) that can be produced in the  hard process and $\mu$
denotes the factorization scale. The coefficient function
$d\hat\sigma_a/dx$ is the differential cross-section for producing
a massless parton $a$ as in Eq.(\ref{facN0}) and with collinear
singularities subtracted in the ${\overline {\rm MS}}$ scheme
(recall that $\Gamma_{ab}$ are universal collinear subtraction
terms that depend only on the QCD time-like splitting functions).
Important feature of Eq.(\ref{fac}) is that power corrections
$(m/Q)^p$, i.e. all terms vanishing in the strict limit $m/Q\to
0$, are neglected. We next explain what is the effect of these
terms and why we may want to neglect them.

There are basically two reasons why such power corrections are
omitted. The first one is that the non-logarithmic terms do not
have universal origin and are therefore not controlled by a
Renormalization Group (RG) equation. They are instead process
dependent and must be calculated perturbatively in any particular
process. The second reason has to do with the size of these terms.
Clearly, when $Q>>m$, such terms will have negligible numerical
effect. Of importance will be only the logarithmic terms that are
resumed with the help of Eq.(\ref{fac}) and terms that are finite
in the limit $m/Q\to 0$ (often referred to as constant term). A
very important implication of the PFF formalism is that while the
constant term cannot be predicted and must be obtained from a
process-dependent, fixed order calculation, it is sufficient to do
that calculation in a massless fashion, i.e. by setting $m=0$ from
the outset. Of course, it is also possible to encounter situation
where the condition $Q>>m$ does not really hold. In such
situations one may want to consider a ``mix" of resummation and
fixed order calculation of the power suppressed term. We will come
back to that point in Section (\ref{SEC4}).

The functions ${D}_{a/\Q}(x,\mu/m)$ in Eq.(\ref{fac}) are the
perturbative fragmentation functions \cite{MN}. They satisfy the
DGLAP evolution equation and can be fully reconstructed from it,
if the initial condition at a scale $\mu = \mu_0$ is known. When
we take $\mu_0 \sim m$ the initial condition ${D}^{\rm ini}_{a}$
cannot contain large logarithms and can be derived from fixed
order perturbative  calculations. Next we describe their
derivation.

\section{DERIVATION OF THE INITIAL CONDITION}

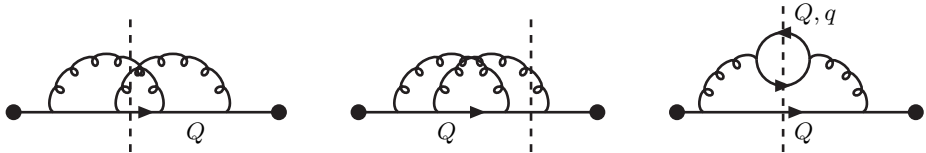
\begin{figure}
\begin{center}
\begin{picture}(200,40)(0,-20)
\SetWidth{1} \Vertex(-70,10){3} \ArrowLine(-70,10)(30,10)
\GlueArc(-35, 10)(20,0, 180){2}{6} \GlueArc(-10,
10)(20,0,180){2}{6} \DashLine(-26,45)(-26,-5){3} \Vertex(30,10){3}
\Vertex(60,10){3} \ArrowLine(60,10)(150,10) \GlueArc(95, 10)(20,0,
180){2}{6} \GlueArc(110, 10)(20,0,180){2}{6}
\DashLine(125,45)(125,-5){3} \Vertex(150,10){3} \Vertex(180,10){3}
\ArrowLine(180,10)(270,10) \GlueArc(230, 10)(20,0, 90){2}{3}
\GlueArc(210, 10)(20,90, 180){2}{3} \ArrowArc(220,30)(10,0,180)
\ArrowArc(220,30)(10,180,360) \DashLine(220,50)(220,-5){3}
\Vertex(270,10){3} \put(-5,0){$Q$} \put(90,0){$Q$}
\put(225,45){$Q,q$} \put(225,0){$Q$}
\end{picture}
\end{center}
\caption{Examples of diagrams that contribute to perturbative
fragmentation of a heavy quark  $Q \to Q + X$ at ${\cal
O}(\alpha_s^2)$. The dashed vertical line indicates the
intermediate (i.e. real emission) state that has to be considered;
$q$ denotes light quark flavor.}
\end{figure}

The most obvious way for the evaluation of the initial condition
for the PFF is by de-convoluting Eq.(\ref{fac}). In such approach
one needs to calculate separately the corresponding differential
distributions for massless and massive quark production to fixed
perturbative order in some process. Historically, such approach
was used in \cite{MN} to derive the initial condition at NLO in
$\alpha_s$:
\begin{equation}
{D}^{\rm ini}_{a/\Q}\left(z,\frac{\mu_0}{m} \right) = \sum_{n=0}
\left({\as(\mu_0)\over 2\pi}\right)^n d^{(n)}_{a} \left
(z,\frac{\mu_0}{m} \right). \label{expansionD}
\end{equation}
with:
\begin{eqnarray}
&&d^{(0)}_a(z) = \delta_{a\Q} \delta(1-z),\nonumber\\
&&d_{a=\Q}^{(1)}\left (z,\frac{\mu_0}{m} \right ) = C_F
\left[{{1+z^2}\over{1-z}} \left(
\ln\left({\mu_{0}^2}\over{m^2(1-z)^2}\right)
-1\right)\right]_+,\nonumber\\
&&d_{a=g}^{(1)}\left (z,\frac{\mu_0}{m} \right ) = T_R \left( z^2
+ (1-z)^2 \right)
\ln\left( {\mu_0^2\over m^2}\right),\nonumber\\
&&d_{a\neq \Q,g}^{(1)}\left (z,\frac{\mu_0}{m} \right )= 0,
\label{DiniNLO}
\end{eqnarray}

Such approach is however too impractical beyond NLO. Next we
describe a better, process independent approach for the derivation
of the initial condition. This approach was first proposed in
\cite{prop} and further elaborated upon in \cite{MM}
where it was also applied for the derivation of all components of
the PFF at order $\alpha_s^2$.

The method is based on the observation that in any process,
collinearly enhanced terms are produced only from diagrams with
real radiation in the external legs. Applying this observation
together with power counting arguments and using the factorization
of phase-space and matrix elements in the collinear limit, one can
derive an explicit expression for the fragmentation function in
Eq.(\ref{fac}). For example, the contributions at NNLO from real
gluon radiation or quark-antiquark pair emission take the form
\cite{MM}:
\begin{eqnarray}
\widetilde{D}(z) = {1\over z} \int [dq_1][dq_2]~W~
\delta\left(1 -z - {(nq_1)\over (pn)} -{(nq_2)\over (pn) }
\right) . \label{Dresult}
\end{eqnarray}

Here, $[dq]$ is the phase space for each of the two emitted
particles with momenta $q_1$ and $q_2$ and $W$ stands for the
appropriately projected square of the matrix element for the
``process" $q(p)\to q+a(q1)+b(q2)$ with the decaying particle
having off-shell momentum $p$. The ``splitting'' function $W$ is
given by $\sim {\rm Tr}\left[\not{\!n} V^{\rm
coll}(p,q_1,q_2,n;m)\right] $ in the case when the decaying parton
is a quark and by $\sim g^{\mu\nu}V^{\rm
coll}_{\mu\nu}(p,q_1,q_2,n;m)$ for decaying gluon. Note that in
the calculation of the matrix $V^{coll}$ we assign to the decaying
particle not the usual spinor or polarization vector but a
propagator as appropriate for particle with momentum $p, ~p^2\neq
m^2$ and mass zero or $m$ depending on its flavor and spin. An
important ingredient to the factorization program is that one
works in a physical gauge defined with the help of an auxiliary
light-like vector $n$. This vector is arbitrary with the only
requirement that its dot product with the hard momentum $p$ is
non-zero i.e. $p.n\neq 0$. The delta function appearing in
Eq.(\ref{Dresult}) simply represents the constraint defining the
observed fraction of the energy. Its argument can be easily
understood in terms of the usual Sudakov parametrization.

Similar expression exists for the virtual corrections as well. In
this case the only change with respect to Eq.(\ref{Dresult}) is in
the argument of the delta function and in the phase-space for real
emission for one of the particles. Examples for the two types of
contributions to the heavy quark $\Q$ initiated component of PFF
are given on Fig.1, and for the gluon initiated component are
shown on Fig.2.

Another unusual feature of this construction (compared to the
usual single particle decay kinematics) is that the contributions
$\sim \delta(1-z)$ that contain the pure virtual corrections arise
from diagrams with a cut through a single line (compared to two
lines in the heavy particle decay case). We do not have to
consider such diagrams in our derivation since all contributions
$\sim \delta(1-z)$ can be completely fixed from the flavor
conservation condition:
\begin{equation}
\int_0^1 dz \left( \D_{\Q/\Q}(z) - \D_{\overline\Q/Q} \right) = 1.
\label{FNC}
\end{equation}

Because the integration range of all integrals covers the full
one- or two-particle real emission phase space, we can exploit for
their evaluation methods and techniques that have become by now
standard. We apply the IBP identities \cite{IBP} constructed from
the vectors $p$ and $n$ and the integration momenta along the
lines of the approach in
\cite{delta}. We apply the Laporta algorithm
\cite{Laporta} implemented in the program AIR
\cite{babis}. We have reduced the problem to the evaluation of
about 40 master integrals. For their evaluation we have made use
of differential equation derived from the IBP identities.

\begin{figure}
\begin{center}
\begin{picture}(200,40)(0,-20)
\SetWidth{1}
\Vertex(-50,10){3}
\Gluon(-50,10)(-20,10){2}{3}
\ArrowArc(10,10)(30,0,180)
\ArrowArc(10,10)(30,180,0)
\Gluon(40,10)(70,10){2}{3}
\DashLine(18,-28)(18,48){3}
\Gluon(-11,-11)(31,31){2}{7}
\Vertex(70,10){3}
\Vertex(130,10){3}
\Gluon(130,10)(160,10){2}{3}
\ArrowArc(190,10)(30,0,180)
\ArrowArc(190,10)(30,180,0)
\Gluon(220,10)(250,10){2}{3}
\DashLine(198,-28)(198,48){3}
\Gluon(169,-11)(169,31){2}{7}
\put(-8,-27){$Q$}
\put(172,-27){$Q$}
\Vertex(250,10){3}
\end{picture}
\end{center}
\caption{Examples of diagrams for the gluon decay process
$g \to Q + X$ at ${\cal O}(\alpha_s^2)$ that contribute to the
function $W$. The dashed vertical line again indicates the
intermediate state.}
\end{figure}
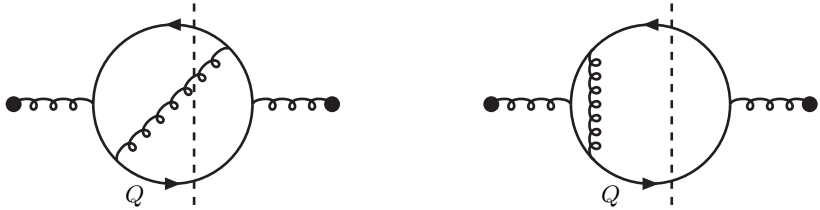
\section{APPLICATIONS}\label{SEC4}

The formalism described in the previous Sections is indispensable
in not-completely inclusive observables where heavy flavored
hadrons (typically mesons) are measured. Such observables are more
complicated and at the same time very interesting because they are
sensitive to the non-perturbative structure of the observed
hadrons. The usual way of treating such processes is the
following: one assumes that a heavy flavored meson is produced at
a scale of the order of the mass of the heavy quark from the
non-perturbative hadronization of a heavy quark with the
corresponding flavor. In hard reactions where the hard scale $Q$
is much larger than the quark's mass one can split the production
of the heavy flavor in a convolution of perturbative and
non-perturbative parts. The perturbative part describes the
production of heavy quark and correctly accounts for all the
radiation at scales down to $\sim m$. Naturally, that function is
correct up to power corrections $\sim (m/Q)^p$. If necessary,
these corrections can be incorporated from a fixed order
calculation (examples are the $p_T$ spectrum of hadrons in
hadronic collisions \cite{KKSS}, \cite{FONLL} and the study of $B$
and $D$ fragmentation in $e^+e^-$ \cite{NO}).

The non-perturbative part of the production of the observed meson
is described by the so-called non-perturbative fragmentation
function $D^{n.p.}(z)$. At the fragmentation stage the fragmenting
quark can radiate only small, of the order of the hadronic scale,
transverse energy. Therefore, it is natural to identify $z$ with
the fraction of the large component of the momentum of the
observed hadron. For example, that can be its longitudinal
momentum fraction $p^{||}_H = z p^{||}_\Q$.

Similarly to the parton distributions, the non-perturbative heavy
quark fragmentation function has to be extracted from experiment.
The cleanest process studied so far is the energy spectrum of
heavy flavors in $e^+e^-$. The most important property of the
function $D^{n.p.}(z)$ is its process independence i.e. it can be
applied to any other process subject only to the restriction
${\cal{O}}(m/Q)^p$.

There are many processes where this formalism can be successfully
applied, like the high $p_T$ spectrum of hadrons in hadron
collisions, photoproduction, $b$-fragmentation in $t$-decay etc.
The later process, for example, presents an interesting method for
the precise determination of the top-quark mass at the LHC. The
knowledge of the initial condition for the PFF through order
${\cal{O}}(\alpha_s^2)$ permits resummation of quasi-collinear
logs with NNLL accuracy. That will, however, be possible only
after the time-like splitting kernels of the DGLAP equation have
been evaluated.

It seems that after the perturbative part has been promoted to the
NNLO/NNLL level, the dominant uncertainty will be associated with
the extracted non-perturbative function. The presently available
data on heavy flavor fragmentation, and particularly on
$b$-fragmentation, from LEP has been already analyzed and no
significant further improvements can be expected (for charm
fragmentation see also \cite{Belle}). Therefore, the viable source
of improvement in the quality of the extracted non-perturbative
component lies with the future International Linear Collider. A
Giga-Z option can supply new data with much higher quality that
will permit new level of precision in the extraction of the
non-perturbative fragmentation function.

\section{SUMMARY}
Production of heavy flavors is becoming an integral part of the
precision physics program of the present and future colliders.
Although the presence of masses leads to theoretical
complications, it is by now well understood how to deal with them
in a systematic fashion. Putting larger effort into the study of
processes of heavy flavor production at colliders will be
rewarding since differential distributions are sensitive to the
non-perturbative hadronization effects.

The successful realization of such program depends crucially on
two factors: improvement of the quality of the perturbative
corrections and precise and systematic extraction and application
of the non-perturbative fragmentation component. On the
perturbative side the future is bright. Many important ingredients
are already known at the NNLO level and the advances in the higher
order perturbative calculations makes plausible the calculation of
all relevant pieces (the most important being the three-loop
time-like splitting functions).

The present state of the non-perturbative components is not as
exciting. Particularly for $b$-fragmentation, the knowledge of
that function is based on analyzes of LEP data which results in
relatively large uncertainty. The situation is somewhat better for
the lowest few moments of the non-perturbative fragmentation
function (for $b$-mesons); the first moment is measured with $\sim
1\%$ accuracy \cite{DELPHI}. However, future prospects for making
significant progress in this direction exist. They are based on
the Giga-Z option of the International Linear Collider. In view of
the impressive state of the perturbative side, improvement in the
quality of the extracted non-perturbative component is mandatory.
Achieving such precision can bring to a qualitatively new level
our understanding of the processes involving heavy flavors. Our
community can make this happen.

\begin{acknowledgments}
It is my pleasure to thank Kirill Melnikov for fruitful
collaboration on this project. Research supported by the US DOE
under contract DE-FG03-94ER-40833 and by the start up funds of the
University of Hawaii.
\end{acknowledgments}

\end{document}